\documentclass[10pt,preprint]{aastex}

\usepackage{graphicx}

\usepackage{natbib}

\newcommand{\be}{\begin{equation}}
\newcommand{\ee}{\end{equation}}
\newcommand{\nn}{\mbox{} \nonumber \\ \mbox{} }
\newcommand{\ba}{\begin{eqnarray}}
\newcommand{\ea}{\end{eqnarray}}

\newcommand{\Alfven}{ Alfv\'{e}n }

\newcommand\etal{\textit{et al.\ }}
\newcommand\eg{\textit{e.g.}}

\newcommand{\Bf}{{magnetic field}}

\newcommand{\BH}{{black hole}}
\newcommand{\ms}{{magnetosphere}}

\newcommand{\Fermi}{{\it Fermi}}

\begin{document}

\title{Resolving Doppler-factor crisis in AGNs: non-steady magnetized outflows.} 

\author{Maxim Lyutikov, Matthew  Lister\\
Department of Physics, Purdue University, \\
 525 Northwestern Avenue,
West Lafayette, IN
47907-2036 }

\begin{abstract}
Magnetically-driven non-stationary acceleration of  jets in AGNs  results in the leading parts of the flow been accelerated to much higher Lorentz factors than in the case of steady state acceleration with  the same parameters. The higher Doppler-boosted parts of the flow may dominate the high energy emission of blazar jets. We suggest  that  highly variable  GeV and  TeV  emission in blazars is produced by the faster moving  leading edges of  highly magnetized non-stationary ejection blobs, while the radio data trace the slower-moving bulk flow.  Model predictions compare favorably with the latest   Fermi $\gamma$-ray and MOJAVE radio VLBI results. 
 \end{abstract}

\maketitle

\section{Introduction}

\subsection{ Bulk Lorentz Factor Crisis in AGNs}

One of the defining characteristics of many AGNs is their flux variability in all spectral bands \citep{Krolik:1999}. 
In particular, extremely short time scales of TeV variability are challenging to the models.  The rapid  flares
reported for Mrk 501 and PKS 2155$-$304, on timescales of 3-5 minutes \citep{2007ApJ...669..862A, 2007ApJ...664L..71A} imply  an emitting size smaller than the gravitational radius $t_{lc}\sim$hours of the supermassive black
holes of these blazars. 
There are two contradictory issues related to short time scale variability. First, it implies  a small emission size, which poses a problem for efficiency of energy conversion into radiation. Secondly, there is the compactness problem \citep{Guilbert:1983}.  If variability is detected in $\gamma$-ray photons of energies exceeding the electron rest mass energy, then the emission region contains photons which can pair produce.  If the number density of these photons is too high, then none of the photons will escape the region. The solution to both  problems is bulk relativistic motion towards the observer, which  reduces the intrinsic luminosity, decreases the implied energy of the photons, and  increases the internal  time scales.   The required  Doppler factor  then exceeds $\delta \geq 100$.

While highly relativistic motion may appear to be a cure-all, in AGNs  the  bulk Lorentz factor $\gamma$ can be directly constrained by VLBI observations of bright blobs moving with apparent speeds on the sky, $\beta_{app}$, that appear to be superluminal.  This type of motion occurs when the emitting region is moving relativistically and close to the line of sight \citep{Rees:1966}.  The apparent transverse motion can exceed $c$ due to propagation effects.  If a blob is moving along with the bulk flow of a jet and its velocity vector makes an angle, $\theta_{ob}$, with the line of sight, then its apparent motion transverse to the line of sight will be:
$
\beta_{app}={\beta \sin{\theta_{ob}}}/({1-\beta \cos{\theta_{ob}}})
$.
The maximum $\beta_{app}$ can reach is $\beta \gamma$ when $\theta_{ob}\cong1/\gamma$.  Thus, if the blob motion corresponds to the underlying bulk motion of the jet, measuring $\beta_{app}$ can constrain the possible bulk Lorentz factor, $\gamma$. 

\subsection{MOJAVE results}

The latest MOJAVE VLBI results do support  the interpretation of moving jet features as physical entities, as opposed to patterns  \citep{2009AJ....138.1874L}. Observations of  bidirectional motions, the near-absence of inward moving features, ejections of  multiple blobs in the same jet  with the same speed, and tight  correlations of jet speeds with other properties, such as  $\gamma$ ray emission, apparent $\gamma$-ray luminosity, brightness temperature,  and even optical classification, all support the notion that the blob motion
reflects the underlying flow.

The MOJAVE survey of compact, highly beamed radio-loud AGN has analyzed the motion of emitting blobs in 127 jets and found that the observed superluminal speed distribution peaks at $\beta_{app}\sim10$ and tapers off at $\beta_{app}\sim 50$ \citep{Lister:2009}.  This suggests that the bulk Lorentz factors of such objects are typically around $\sim10$, and extend up $\sim50$, making the estimated values of $\gamma \geq 50$ for PKS 2155$-$304 and Mrk 501  rather difficult to reconcile with the radio data.  Furthermore, direct VLBI observations of these sources on parsec scales have not even detected superluminal motion \citep{Piner:2004,Giroletti:2004}.

VLBI observations of blazars such as PKS 2155$-$304 and Mrk 501 are not the only data which imply a low $\gamma$.  Another way of investigating blazars is to search for their AGN counterparts whose jets are not directed along the line of sight, which are presumed to be radio galaxies (there are actually two distinct types of radio galaxies, FRI and FRII, that are thought to correspond to the two categories of blazars, BL Lacs and optically violent variable quasars \citealt*{Urry:1995}.)  However, studies comparing the relative fluxes and numbers of radio galaxies and blazars point towards Lorentz factors of $\gamma <10$ \citep{Henri:2006}. Indeed, preliminary results of MOJAVE observations of low luminosity jets from radio galaxies imply speeds of $c$ or less. 
Thus, there is apparent contradiction between measured superluminal velocities and bulk Lorentz factors required by radiation modeling. This is known as  the  {\it "Blazars'  Bulk Lorentz Factor Crisis"} \citep{Henri:2006}.
 
   \section{Non-stationarity in AGN  flows}

Following \cite{BlandfordRees}, models  of AGN jets typically assume steady-state injection conditions. This is based on the fact that the sonic time over the \BH\ horizon is typically shorter than any observed times scales of variability of \BH\ systems (except for such subtle effects as quasi-periodic oscillations in Galactic binaries, \eg, \citealt*{2005AN....326..798V}). By consequence, it is argued, a system reaches an quasi-equilibrium state. On the other hand, both Galactic \BH\ candidates, as well as AGN jets show a wide variety of non-stationary behavior. This non-stationarity is driven by various  disk instabilities, occurring  on the viscous time scale of (inner) accretion disk.

 The efficiency of the BH-powered jet depends (\eg,\ in a \cite{BlandfordZnajek}  paradigm of jet launching) both on the parameters of the \BH\ (mass and spin) as well as on the \Bf\ supplied by the disk. In addition, as we argue in this paper, {\it jet acceleration may proceed more efficiently in case of non-stationary outflows}    \citep[see also Lyutikov 2010, submitted, ][]{2010arXiv1004.0959G}.
 
  One expects that accretion onto the \BH\ may change on the  time scale of the order of viscous time scale of the inner part of the disk, which may be weeks to months  to years for a typical AGN (\eg, if we associate blob ejection time scales as being due to the inner disk instability). As the relativistic jet propagates away from the central \BH\ it expands sideways, reaching a transverse scale of a fraction of parsec at the distance of the order of a parsec. 
    Close to the \BH, the corona of the accretion disk  is hot, with a nearly relativistic speed of sound:  the sonic time scale  of the \ms\ across the \BH\ is much shorter  than the
  viscous time scale of the inner accretion disk. Further out, the variability time scale of the jet will be shorter than the dynamical time scale of the corona across the jet. 
  Observationally, the connection of the disk and jet activities is  well established in the case of Galactic microquasars \citep{1999ARA&A..37..409M}. In AGN, the data are controversial \citep{2009ApJ...704.1689C}. 

  Let us assume that a  jet with  an opening angle of $\Theta_j \sim 0.1$
is propagating through
  a  corona, which is  in hydrostatic equilibrium with the  
local  sound speed  $c_s$  close to the virial velocity, $c_s =  \sqrt{G M_{BH} /r}$.  Relating the disk variability time scale to the Keplerian period near the \BH, $t_d = \xi r_{BH} /c \approx  10\,  {\rm days}\,  M_{\odot, 9}\,   \xi_2  $, $\xi \gg 1$, where the sound crossing time across  the jet  is longer than the variability time scale for
\be
r_{\rm breakout} \geq  \left( { \xi \over \Theta_j} \right)^{2/3}  r_{BH} =  2 \times 10^{16} M_{\odot, 9}  \xi_2  ^{2/3} \Theta_{j,-1} ^{-2/3} \, {\rm cm}. 
\label{r}
\ee
where the notation $X_a $ implies $(X/10^a)$  (\eg, $ \xi_2= \xi/100$, $\Theta_{j,-1}= \Theta_{j}/0.1$). 
At distances larger than (\ref{r}), the jet variability on a time scale of a hundred Keplerian periods near the \BH, $\xi_2=1$ proceeds on time scales  shorter than the 
dynamical time scale across the jet, so the external medium does not have time to react to changing jet conditions. It is also required that the jet travel time to $r_{\rm breakout}$ be smaller than the the sound crossing time at $r_{\rm breakout}$. This requires
$\xi < 1/ \Theta_j^2$.

 A newly created jet  will then propagate along a  nearly  empty channel, cleared by the previous jet activity. Thus, we expect that  initially,    close to the \BH,  the non-stationary injected jet  propagates through a relaxed corona and will have to ''bore'' its way through. After reaching the distance (\ref{r}), 
 the leading edge of the jet will break out into low density medium created by the previous jet activity.
 What are the consequences of this  non-stationary behavior for jet acceleration?

As a model problem, we   
  assume that a period of accretion 
  lasting time $t_{\rm d}$ has 
  brought onto the \BH\ the \Bf, initiating the Blandford-Znajek process. 
  Since the external medium has fairly low density, the jet expansion is relativistic (see Eq. \ref{gammaj}). 
  In a time $t_{\rm d}$ 
  the jet  inflates a bubble of linear size $c t_{\rm d} \sim r_{\rm breakout}$.   At a distance $ r_{\rm breakout}$, where  the sonic time through the \ms\ is of the order of the jet variability time scale, the newly inflated bubble reaches a near vacuum. What is the behavior of the jet starting at this point?
  
  As a model problem we considered 
  (Lyutikov, submitted)
  a one-dimensional flow
of cold magnetized   into a vacuum and into an external medium of density $\rho_{\rm ex}$. This is reviewed in \S \ref{Riemann}.
In \S  \ref{imply} we apply the results to non-stationary outflows in AGNs. 
 
 \section{Non-stationary relativistic expansion}
\label{Riemann}

            \subsection{Riemann problem for relativistic expansion of magnetized gas}

      Let us assume that before the breakout  into the low density medium the  jet  plasma is moving with velocity $\beta_w$ towards the external medium.        We  have found 
  (Lyutikov, submitted) 
  an    exact self-similar  solution of the relativistic Riemann problem for the  expansion  of  cold plasma with density $\rho_{0}$ and \Bf\ $B_0$ (magnetization parameter $\sigma=B_0^2/\rho_{0}$; the \Bf\ is normalized by $\sqrt{4 \pi} $),   moving  initially with velocity $v_w$ towards the vacuum interface
 \ba &&
\delta_\beta = \delta_\eta^{2/3} \delta_{A,0}^{2/3} \delta_w ^{1/3}
\nn &&
\delta_A ={ \delta_{A,0}^{2/3}  \delta_w ^{1/3} \over \delta_\eta^{1/3}},
\label{main}
\ea
where  the Doppler factors $\delta _a= \sqrt{(1+\beta_a)/ ( 1-\beta_a)} $  are defined in terms of  the plasma velocity $\beta$, local \Alfven velocity $ \beta_{A}$, self-similar parameter $\eta= z/t$, initial wind velocity $\beta_w$ and the 
   \Alfven velocity in the undisturbed plasma $ \beta_{A,0}$.
These equations give the velocity $\beta$, density $\rho = U_A^2 \rho_0/\sigma$ (where $U_A= \beta_{A} /\sqrt{1-\beta_{A} ^2}$) and proper  \Bf, $B =(\rho/\rho_0) B_0$ as a function of the self-similar variable $\eta= z/t$ (the expansion of plasma  starts at $t=0, z=0$ and proceeds into a positive direction $z>0$).  We stress that these solutions are exact, no assumptions about the value of the parameter $\sigma$ and velocity $v_w$ were made.

\begin{figure}[h!]
 \begin{center}
\includegraphics[width=.49\linewidth]{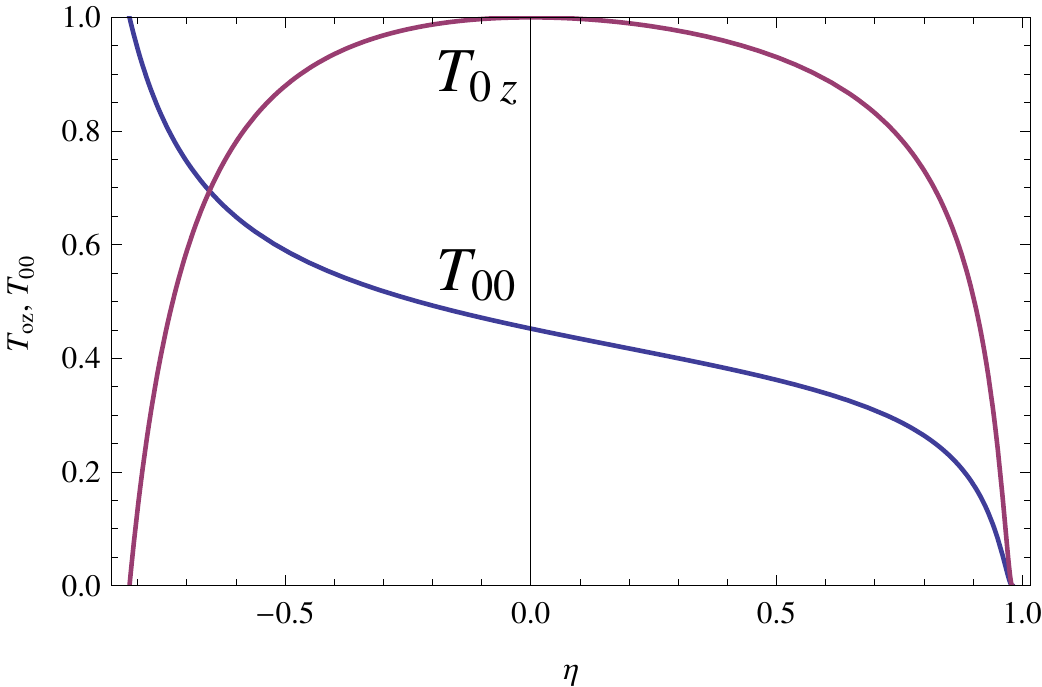}
\includegraphics[width=.49\linewidth]{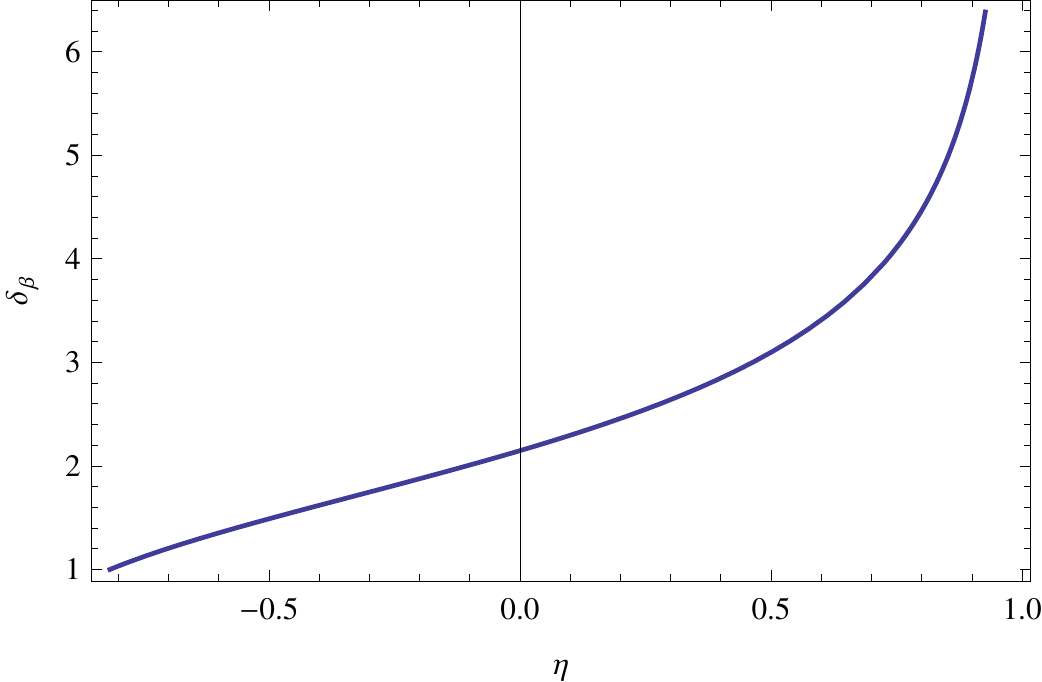} 
\end{center}
\caption{ Energy density $T_{00}$ and energy flux $T_{0z}$  ({\it Left panel}) and Doppler factor $\delta_\beta = \sqrt{(1+\beta)/(1-\beta)} $ ({\it Right panel}) for self-similar expansion into a vacuum as functions of the self-similar coordinate $\eta = z/t$. Stationary initial conditions are assumed.  The energy density is normalized to values in the undisturbed medium, while the energy flux is normalized to maximum values at $\eta = 0$. Solutions extend from the front of the rarefaction wave at 
$\eta_{RW} = - \sqrt{\sigma/(1+\sigma)}$ up to the vacuum interface $ \eta_{\rm vac} =  2 { \sqrt{\sigma(1+\sigma)}\over (1+2 \sigma) }  $. The Doppler factor is unity in the undisturbed plasma and reaches a value corresponding to $\gamma_{vac} = 1 + 2 \sigma$ at the vacuum interface. Plots are for $\sigma =2$. }
\label{T00T0z}
\end{figure}

Particularly simple relations are obtained for plasma  initially at rest expanding into vacuum  $\beta_w=0, \,  \delta _\beta=1$ 
(Lyutikov, submitted). 
   The flow accelerates  from rest towards the vacuum interface (Fig. \ref{T00T0z}).
The bulk of the flow is moving with Lorentz factor $\gamma' \sim \sigma^{1/3} $. The flow becomes supersonic  at $\eta =0$, at which point 
      $\gamma'= (\sigma/2)^{1/3}$.    The  vacuum interface moves  with Lorentz factor $\gamma_{vac} '= 1+2 \sigma$.
      In the observer frame the vacuum interface is moving with $\delta _\eta=  \delta_{A,0}^{2} \delta_w$, which 
       in the limit $\sigma, \, \gamma_w \gg1$ gives
\be
\gamma_{vac} = 4 \gamma_w  \sigma.
\label{gammavac}
\ee
   As the flow expands,  
      the energy flux $T_{00}$ and the energy density $T_{0z}$  stay nearly constant in the bulk of the flow at a value $\approx B_0^2/4$.
      The energy flux is maximal at the sonic point $\eta =0$, while the energy density slowly decreases towards the vacuum interface (Fig. \ref{T00T0z}).
   
 Most importantly,   in a narrow region near $\eta_{\rm vac}   =  2 { \sqrt{\sigma(1+\sigma)}/ (1+2 \sigma) }   \approx  1- 1/( 8 \sigma^2)$,  
 the Doppler factor increases from $\sim \sigma^{1/3}$ in the bulk to  $\delta_{\beta, max} \sim 4  \sigma$ on the vacuum interface. Values of 
 $\delta_{\beta} > (1/2) \delta_{\beta, max}$ are reached within a range of $\Delta \eta = 7/(8 \sigma^2)$ near the vacuum interface.  The relative  amount of energy with 
 Doppler factors $\delta_{\beta} > (1/2) \delta_{\beta, max}$ is $\sim 3/\sigma^2$ (the total energy of the outflow at time $t$ is
 $E_{tot}  =  \rho_0 (1+\sigma/2) \sqrt{\sigma \over 1+ \sigma } t \approx  B_0^2 t/ 2$). 
  In addition, in a non-self-similar regime,  most of plasma will reach the
     vacuum terminal velocity, since the forward characteristics never cross the vacuum interface regardless of the dimensionality of the flow 
     \citep[][see also Lyutikov, submitted]{GreenspanButler,ZeldovichRaizer} .
     
\begin{figure}[h!]
 \begin{center}
\includegraphics[width=.49\linewidth]{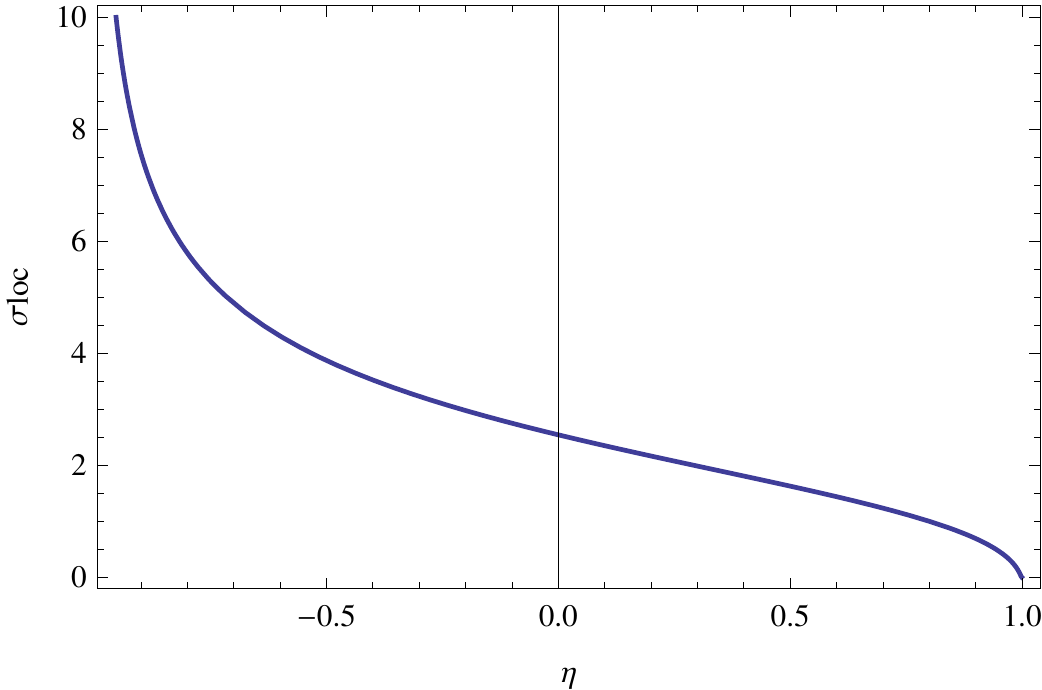}
   \includegraphics[width=.49\linewidth]{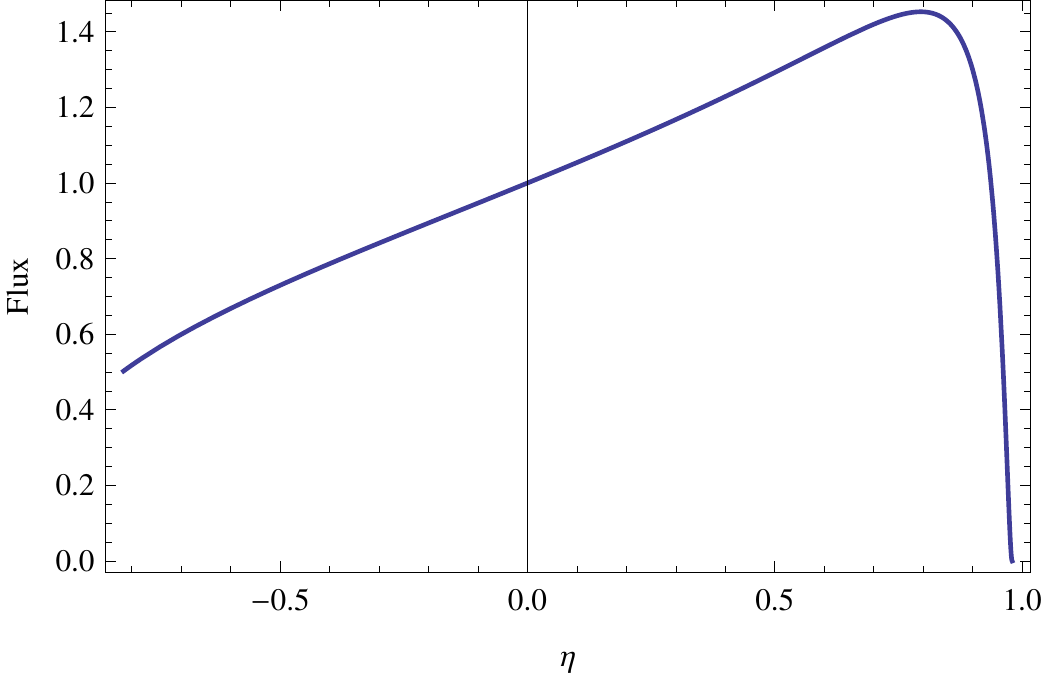}
\end{center}
\caption{({\it Left panel}) Local magnetization $\sigma_{loc}$, as given by Eq.  (\ref{sigmaloc1}).  Initially $\sigma =10$. {\it Right panel}:  Observed flux produced by the plasma expansion into vacuum, parametrized as a product of the rest frame energy density and the Doppler factor cubed.  The scale is normalized to the value at $\eta=0$.}
\label{sigmaloc}
\end{figure}

     As the flow expands, the local magnetization 
     \be
     \sigma_{loc} = {  B^2 \over \rho} \left({ \delta_{A,0}^{2/3}  \over  \delta_\eta^{1/3}} - { \delta_\eta^{1/3} \over  \delta_{A,0}^{2/3} }\right) 
     \label{sigmaloc1}
     \ee
     decreases (Fig. \ref{sigmaloc}).
     At the sonic point $   \sigma_{loc}  = (\sigma/2)^{2/3}$.

\subsection{Expansion into medium}

If there is an outside medium with density $ \rho_{\rm ex}$, we may identify two expansion regimes with different   properties of the forward shock. For relativistically  strong forward shocks, so that the post-shock pressure is much larger than density,  the Lorentz factor of the contact discontinuity (CD) is 
 \be
\gamma_{CD} =
\left({3 B_0^2 \gamma_w^2 \over 8 \rho_{\rm ex}} \right)^{1/4}  \approx \left({ L\over \rho_{\rm ex} c^3 } \right)^{1/4}  r^{-1/2}
\label{gammaCD}
\ee
(the last approximation assumes $\sigma \gg 1$). Here $L \approx   B_0^2 \gamma_w^2 c^3 r^2$ is the jet luminosity.
For weak forward shocks,  the velocity of the CD approaches the expansion velocity into vacuum $\gamma_{vac}$ (Eq. \ref{gammavac}). The transition between the strong  and weak shock occurs  for 
\be
\sigma_{crit} = \left({ 3 \over 2048 \gamma_w^2} {\rho_0 \over  \rho_{\rm ex}} \right)^{1/3}.
\label{rhoex1}
\ee
For $\sigma < \sigma_{crit}$, the forward shock is weak.
In the limit $\sigma  \gg 1$  the vacuum approximation is applicable for 
    \be
n_{ex,0}= {     \rho_{ex,0}  \over m_p} \leq { 3 L \over 512 \pi \gamma_w^4 \sigma^4 c^3 r^2} = 
  5 \times   10^{-8} \, {{\rm cm^{-3}}}L_{46} \left( {r \over  0.1 {\rm pc}} \right)^{-2} 
   \times \left( {\gamma_w  \over  10} \right)^{-4} \left( {\sigma  \over  10} \right)^{-4}, 
   \label{11}
    \ee
    where numerical estimates are given for typical parameters in AGNs.
  The required density is fairly low, yet it depends sensitively on the parameters of the flow. Even if the density is higher than given by Eq. (\ref{11}), the resulting forward shock will propagate with a Lorentz factor that is only weakly dependent on the external density: 
    \be
    \gamma_{FS} \sim 2 \gamma_w \sigma 
    (\rho_{ex} /\rho_{ex,0} )^{-1/4}= 200  \times \left( {\gamma_w  \over  10} \right)^{-4} \left( {\sigma  \over  10} \right)^{-4}  (\rho_{ex} /\rho_{ex,0} )^{-1/4}. 
    \ee

   \section{Implications for non-stationary acceleration in AGNs}          
   \label{imply}

        We propose that flares in  TeV and GeV emitting  blazars are produced in the leading expansion edges of non-stationary ejection events,  moving with $ \gamma _{\rm max }\geq 100$, while the observed velocities of the radio blobs  correspond to the bulk motion with $\gamma_{\rm bulk}\sim 10-50$.  In this section we discuss observational implications of the model.

       In our picture,  after the switch-on of the acceleration process, the AGN central engine  produces a relativistic jet, and as the latter bores through the corona, it can reach a relativistic  Lorentz factor $\gamma_w$.
    The value of    $\gamma_w$ can be determined by the pressure balance (Eq. \ref{gammaCD}) with high external density:
    \be
    \gamma_w = \left({ L\over \rho_{\rm ex} c^3 } \right)^{1/4}  r^{-1/2}  = 20 L_{j, 46}^{1/4} \Theta_{j,-1} n^{-1/4},
\label{gammaj}
    \ee
    where we have assumed $\xi = 1/\Theta_j$ and typical parameters of AGN outflows. 
       
After breakout, at  $r> r_{\rm break}$,   the total Lorentz factor in the observer frame will be $2  \sigma ^{1/3}$ times larger for the bulk flow, $  \sim 2 \gamma_w\sigma ^{1/3}$,   and $4 \sigma$ times larger for the leading edge,
       $ \sim 4  \gamma_w \sigma $:
       \ba && 
       \gamma_{\rm bulk} = 40 \,  \sigma^{1/3} \,
       L_{j, 46}^{1/4} \Theta_{j,-1} n^{-1/4} 
       \nn &&
        \gamma_{\rm vac} =  80 \,  \sigma
\,  L_{j, 46}^{1/4} \Theta_{j,-1} n^{-1/4} .
\label{12}
        \ea
       For values of $\sigma$ exceeding unity, the difference between Lorentz factors of the bulk flow and that of the leading edge is even greater.

Soon after the break out,  in the self-similar stage, the relative amount of energy in the fast leading tail is fairly small in the case of  highly magnetized jets,  $\sim 3/ \sigma^2$, yet it may dominate the beamed high energy emission due to high Doppler boosting. For example, 
if we parametrize the observed intensity produced by the jet as a product of the rest-frame energy density, $  T_{0z}/(\gamma^2 \beta)$,  and Doppler factor cubed $\delta_\beta^3$, it will be dominated by the fast moving parts of the flow (Fig. \ref{sigmaloc}).

In addition, at later stages of expansion of  a finite size blob, a larger fraction   of the material may be accelerated to the maximal Lorentz factor. If the initial blob had a size $L$ in its rest frame, most of the blob's material gets accelerated to Lorentz factor $
\gamma_{\rm bulk}' \sim  1+ 2 \sigma $  in time $\Delta t \sim 8 \sigma^2 L/c$  
(Lyutikov, submitted). For a blob moving with $\gamma_w$ the acceleration  time in the observer frame is further extended to
$\Delta t \sim 16 \sigma^2 \gamma_w L/c $. This time is, typically, longer than the time scale for blob interaction, Eq. (\ref{rmerge}). Thus, the bulk of the material does not get accelerated to the maximum Lorentz factor  $ \gamma _{\rm max }$. 
   
   There are a number of correlations that we would expect in the model. 
First of all, we expect some correlation between the $\gamma$-ray and radio core fluxes, and this has already been seen by Fermi and MOJAVE  \citep{2009ApJ...696L..17K}.

 Since the $\gamma$-rays  are produced in a faster moving part of the flow, we expect that 
  the  jets of $\gamma$-ray selected  AGNs are more aligned than those in radio-selected
  samples.  This effect should be reflected in a flux-flux plot of a well-defined flux-limited blazar sample, due to
the different degrees of Doppler beaming in the radio versus $\gamma$-ray regimes. For example, 
a tight linear correlation of intrinsic (i.e., unbeamed) radio vs.  $\gamma$-ray jet luminosity will 
be greatly smeared out in the flux-flux plot,  although an upper envelope will still be present (see, e.g., simulations by \citealt*{2007AIPC..921..345L}). 
Indeed, the form and scatter of the observed radio-$\gamma$-ray  correlation  in blazars \citep{2009ApJ...696L..17K}  does support the notion that the $\gamma$-ray emission is likely boosted by a higher Doppler factor than the radio emission.   This is also reflected in the superluminal speeds and apparent jet opening angles, which suggest higher Doppler factors and smaller viewing angles for Fermi-detected AGN \citep{2009ApJ...696L..22L,2009A&A...507L..33P}  

On the other hand, a radio-$\gamma$-correlation depends crucially on the fact that 
the cores of   AGN jets are optically thick to synchrotron emission up to the frequency-dependent  radius  
 \citep[][Eq. 28]{Blandford:1979}
  \be
  r_{\rm core} \approx 1.4 {\rm pc} \zeta_R^{2/3} L_{46}^{2/3} \gamma_{w,1}^{-1/3} \nu_9^{-1} .
  \label{core}
  \ee
  where $\zeta_R$  parametrizes the  observed radio luminosity in terms of the total jet power $L_R =\zeta_R L$, with $\zeta_ R \sim 0.01$;
  we also assumed here (and in the estimates below) that $\Theta_j \sim 1/\gamma_j$, $\xi = 1/\Theta_j^2$. 
The  core radius  (\ref{core}) is typically much larger than the breakout radius (\ref{r}).  Thus, typically, the jet breakout will occur while the jet is still optically thick in the radio. In this case the high energy emission will not be accompanied by the simultaneous increase of radio flux (at least at low frequencies). At sufficiently high radio frequencies,  the radius where the jet becomes optically thin (\ref{core}) may become comparable to the  breakout radius (\ref{r}).

   We expect that {\it  $\gamma$-ray event should precede the radio blob ejection} by 
  \be
  \Delta t_{\gamma-R} \sim { r_{\rm core}/c \over 2 \gamma_w^2} \sim 8  \, \zeta_R^{2/3} L_{46}^{2/3} \gamma_w^{-7/3} \nu_9^{-1} \, {\rm days}.
  \label{tgamma}
  \ee 
  These are typical time scales between variability at  radio, optical and $\gamma$-ray bands \citep{2010ApJ...710L.126M}. 
  Also, since this timescale (Eq. \ref{tgamma}) is fairly long, we expect occasional  $\gamma$-ray flares without radio blob ejection; this appears to be the case in TeV flares associated with high-energy peaked AGN such as Mrk 421, and Mrk 501, in which no major changes in VLBI radio structure are seen (MOJAVE program; Ros \etal\ in prep).  
  Furthermore, the  {\it $\gamma$-ray events   should be  better correlated with with radio blob ejection  at high radio frequencies.}  Higher resolution monitoring of MOJAVE sources by the Boston University group \citep{2005AJ....130.1418J} at 43 GHz has revealed instances of jet features moving at higher speeds than those seen at 15 GHz \citep{2009AJ....138.1874L} that fade out very rapidly close to the core region. The extensive VLBA+Fermi data currently being gathered on these AGN should provide a useful test of this prediction.

  The above correlations assume that the $\gamma$-ray photons produced at  the breakout radius (\ref{r}) do not suffer from absorption. This depends on the compactness parameter corresponding to the bulk motion of the jet (since $\gamma$-ray photons can pair produce on a lower moving bulk plasma):
  \be
   l \sim { 1 \over \gamma_w^6} { \sigma_T \over m_e c^3} {L_\gamma \over r_{\rm breakout}} = 10^{-7} L_{\gamma, 44} \Theta_{j,-1}^8
  \ee
  On the other hand, the $\gamma$-ray  variability time scale depends on the Lorentz factor of the leading edge
  \be
  \Delta t  \sim {r_{\rm breakout} \over c} {1 \over 32 \gamma_j ^2 \sigma ^2} = 150\, {\rm sec} \, {1\over \sigma^2} \,  \Theta_{j,-1}
  \ee
  Thus,   {\it the model is able to accommodate  both the  requirement of small optical depth for $\gamma$-ray photons  and the short time-scale variably, down to few minutes. } 
  
\begin{figure}[h!]
 \begin{center}
   \includegraphics[width=.99\linewidth]{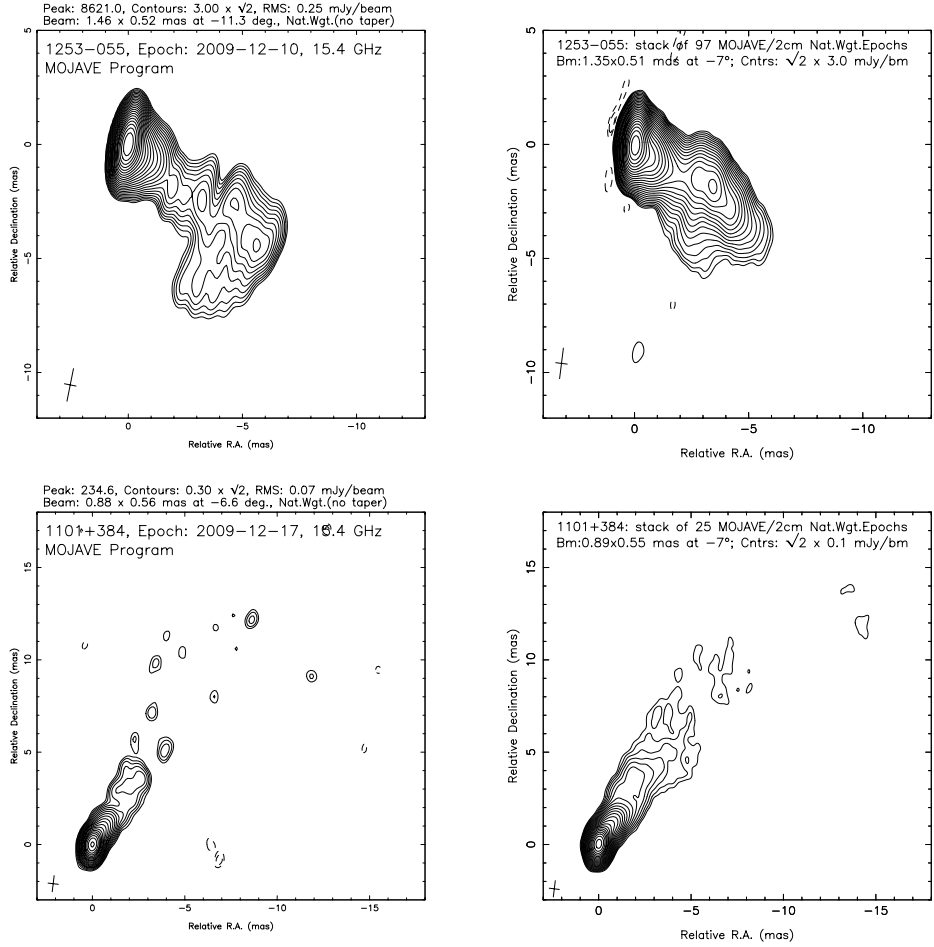}
\end{center}
\vskip -3 truein
\caption{MOJAVE 15 GHz VLBA images of low-energy-peaked quasar 3C 279 (1253$-$055; top two panels), and high-energy-peaked BL Lac object Mrk 421 (1101$+$384; bottom two panels). The left-hand panels show the parsec-scale jet structure in a recent MOJAVE epoch. The right-hand panels show time-averaged images created by combining archival VLBA epochs from 1995 to 2009. The relatively smooth fall-off in jet intensity downstream from the bright core in Mrk 421 is a characteristic of high-energy-peaked jets in the MOJAVE sample. These jets display very few changes in their radio jet structure over time, as indicated by the similarity in the inner 5 milliarcseconds of the single epoch and stacked images of Mrk 421. By contrast, the jets of flat-spectrum radio quasars such as 3C 279 generally display a constantly varying structure that is dominated by numerous bright superluminal knots \citep{2009AJ....137.3718L,2009AJ....138.1874L}.}
\label{knotty-smooth}
\end{figure}

The jets of lower power FRI sources and higher power FRII sources have somewhat different morphology on mas-scales.  Relatively low power blazar jets in the MOJAVE program like Mrk 421 or Mrk 501  show a very smooth fall-off in radio intensity with distance from the core, and have maintained this structure over more than a decade of VLBI monitoring. In contrast, higher power blazars such as 3C~279 display jet morphologies dominated by individual bright knots that continually emerge from the core and move downstream at superluminal speeds (see Fig. \ref{knotty-smooth}).

One possible explanation for the dichotomy of jet properties is that the emitted blobs merge, creating smooth large scale profiles. 
  If the blobs are ejected on a time scale $t_d = \xi r_{BH} /c$, they will merge at distance (in the observer frame the time of merger is determined by the 
  Lorentz factor of the trailing edge of the preceding injection $\gamma \sim \gamma_j/(4 \sigma)$:
  \be
  r_{\rm merge} \sim 2 c t_d \left( {\gamma_j ^2 \over 16 \sigma^2} \right) \approx 3 \times 10^{20} {\rm cm} \, {1\over \sigma^2} \,  \Theta_{j,-1}^{-2}
\label{rmerge}
  \ee
  Thus, at linear scales $\geq 100$ pc, the jet is expected to be mostly smooth. Note, that since in blazars the angle between the jet direction and the line of sight are small, the projected distance corresponding to (\ref{rmerge}) are small.  For example, with the jets in Mrk 421 or Mrk 501  at z=0.03 oriented at a few degrees to the line of sight, this corresponds to a few milliarcseconds projected on the sky. 
In addition, 
since the radius $ r_{\rm merge}$ depends sensitively on the assumed bulk Lorentz factor, our model predicts that in powerful FR-II jets,  the jet may remain knotty up to the kiloparsec scale. 

Can the observed morphology of the jets be used to determine intrinsic jet properties, like bulk Lorentz factor and magnetization? High-frequency peaked blazars (HBLs) are under-represented in the MOJAVE sample; in fact, there are no HBLs in the complete radio-selected sample \citep{2005AJ....130.1389L}. On the other hand, HBLs are well represented in the \Fermi\ sample \citealt{2009arXiv0912.2040A}. In the framework of our model, this can be due to their lower bulk Lorentz factors, but high magnetization, see Eq. (10). It would then imply relatively smaller merging distances (15) and smoother jet structures, consistent with observations (Fig. 3). An obvious caveat in this argument is that HBLs have flatter $\gamma$-ray spectra and are more likely to be seen by the \Fermi\ LAT. Further careful analysis of beaming and instrumental selection effects is needed to explore these possibilities more fully. 

   FRI and FRII  sources could also have different intrinsic variability times scales. In our present model, we relate the variability times scale to the mass of the central black hole, which does not vary considerably between FRI and FRII sources 
\citep{2001A&A...379L...1G}. 

 \section{Discussion}

 In this paper 
we discuss  the effects associated with non-stationarity of the jet ejection. In particular, we argue that  the leading edge of a
non-stationary {\it magnetized} outflow  can achieve a bulk Lorentz factor much larger than
 would be inferred for a steady state outflow given similar conditions. In the case of the expansion of a highly magnetized plasma, the ratio  of Lorentz factors of  the bulk flow and that of the leading edge can be as high as $2 \sigma^{2/3}$ ($\sigma$ is plasma magnetization). This ratio can reach tens for highly magnetized flows with $\sigma \sim 10$. We suggest that the Doppler factor crisis in AGNs (a difference of the Doppler factors  inferred from radiation modeling, especially of short time scale TeV flares and from the observations of radio blobs) may be resolved by non-stationary outflows: 
highly variable emission is produced by the fast-moving leading edge of an expansion, while the radio data trace the slower-moving bulk flow.  

The suggested model is qualitatively different from the internal shock models, where non-stationarity is invoked to produce shocks and dissipate the energy of the relative bulk motion of the colliding media. Collision of  strongly magnetized plasma blobs results in  only weakly dissipative  internal shocks. The fast leading expansion edges will  generate powerful  shocks in the surrounding medium that may produce the high energy emission. We do not address the question of how magnetic and bulk energy is converted into radiation.

A somewhat similar {\it continuous} acceleration mechanism was proposed by \cite{Tchekhovskoy} \citep[see also][]{2009arXiv0912.0845K}.  It relies on {\it sideways} expansion of the jet after the  break out. Sideways expansion of unconfined magnetically dominated plasma proceeds with Lorentz factor $1+2\sigma$, so that the total Lorentz factor (of a plasma near the edge, affected by the rarefaction wave) is $(1+2\sigma) \gamma_w$, two times smaller than for the case of expansion wave propagating along the direction of motion.  This factor of 2 may have an important effect on escape of high energy radiation, since the optical depth to pair production scales approximately as $\Gamma^{-6 }$, a difference in a factor of 2 in  $\Gamma $ will result in a difference of $64$ in the optical depth.   (\cite{Tchekhovskoy} cannot treat  parallel acceleration akin to breaking into vacuum since the rotation of the central engine is turned on gradually for numerical reasons.)

In summary, we believe that magnetically-driven non-stationary jet acceleration can provide a potential resolution of the longstanding bulk Lorentz factor crisis in blazars. The rich Fermi-VLBA dataset that is currently being gathered on a broad set of AGN should provide an excellent means of testing our proposed scenario. 

Acknowledgments:  This research has made use of data from the MOJAVE database that is maintained by the MOJAVE team. The MOJAVE program is supported under National Science Foundation grant 0807860-AST and NASA-Fermi grant NNX08AV67G. The National Radio Astronomy Observatory is a facility of the National Science Foundation operated under cooperative agreement by Associated Universities, Inc. We would like to thank Roger Romani for discussions.

\bibliographystyle{apj}
\bibliography{/Users/maxim/Home/Research/BibTex}

      \end{document}